# Solvation and Transport of Lithium Ions in Deep Eutectic Solvents

H. Srinivasan[1], V.K. Sharma[1], R. Mukhopadhyay[1,2] and S. Mitra[1,2*]

[1]Solid State Physics Division, Bhabha Atomic Research Centre, Mumbai 400085, India

[2]Homi Bhabha National Institute, Anushaktinagar, Mumbai 400094, India

* Corresponding author: smitra@barc.gov.in

Tel:+91-22-25594674; FAX:+91-22-25505151

# Abstract

Lithium based deep eutectic solvents (DESs) are excellent candidates for eco-friendly electrolytes in lithium ion batteries. While some of these DES have shown promising results, a clear mechanism of lithium ion transport in DESs is not yet established. This work reports the study on the solvation and transport of lithium in a DES made from lithium perchlorate and acetamide using Molecular Dynamics (MD) simulation and neutron scattering techniques. Based on hydrogen bonding (H-bonding) of acetamide with neighbouring molecules/ions, two states are largely prevalent: 1) acetamide molecules which are H-bonded to lithium ions (~ 36 %) and 2) acetamide molecules that are entirely free (~ 58%). Analysing their stochastic dynamics independently, it is observed that the long-range diffusion of the former is significantly slower than the latter one. This is also validated from the neutron scattering experiment on the same DES system. Further, the analysis the lithium dynamics shows that the diffusion of acetamide molecules in the first category is strongly coupled to that of lithium ions. On an average the lithium ions are H-bonded to ~ 3.2 acetamide molecules in their first solvation. These observations are further bolstered through the analysis of the H-bond correlation function between acetamide and lithium ions, which show that ~ 90% of lithium ionic transport is achieved by vehicular motion where the ions diffuse along with its first solvation shell. The findings of this work are an important advancement in understanding solvation and transport of lithium ion in DES.

# 1. Introduction

Deep eutectic solvents (DESs) have become ubiquitous in a variety of applications ranging from manufacturing to pharmaceutical industries. Since their discovery[1-4], DESs have found tremendous applications in the field of electrochemistry – electrodeposition of metals[5-7], lithium ion battery electrolytes[8-10], electroplating[11, 12], supercapacitors[13] etc. These solvents are found to posses properties very similar to ionic liquids, but have an advantage of lower toxicity and better biodegradability. Generally, DESs are mixtures of salt with hydrogen bond donor (HBD) molecules in a particular molar ratio, corresponding to their eutectic point[3, 4]. The freezing point of the mixture is significantly lower compared to their parent compounds. The strong hydrogen bonding between the species of the mixture is suggested to be the key reason for the large depression in their freezing points[3].

A number of deep eutectic solvents (DESs) based on lithium salts have been synthesised[8-10, 13-15]. Recently, such DESs with N-methylacetamide as the HBD have been found to be good candidates for lithium ion battery electrolytes[10]. Especially, it is found that the electrolytes show a passivating behaviour towards the aluminium collector surface, suggesting it is less corrosive compared to the regular solvents like ethylene carbonate[10]. Electrochemical double layer capacitors were also developed using electrolytes based on DES of acetamide and lithium salts like LiClO4, LiTFSI etc[13]. They were found to be very efficient and stable with activated carbon electrodes and the DES as electrolytes[13]. While there have been various studies on the structure[9] and electrochemical[8, 14, 16] properties of such solvents, there have been no detailed investigation on the diffusion and transport of lithium ions in the DESs. Understanding the transport of lithium ions in the solvent is the key to improve their properties for electrolytic applications, in particular as electrolytes for lithium ion batteries. In fact, the observed diffusivities can be useful to describe conductance mechanisms in modified Debye-Hückel theory for salt-organic solvent electrolytes[14].

Ionic liquid (IL) based electrolytes for lithium ion batteries have been explored quite extensively as potential substitutes to combustible organic solvents[17, 18]. However, some major impediments for lithium transport in IL based electrolytes are large sizes of the ions, strong Coulumbic interaction and chemical instability. In the case of DES based electrolytes, it is possible to eliminate the first two drawbacks by using small and neutral HBDs. Considering the nature of solvation of lithium ions in IL electrolytes, two principle mechanisms of its diffusion have been suggested – vehicular motion and structure diffusion[18, 19]. In the case of the vehicular motion, the lithium ion diffuses along with its first solvation shell, i.e. the entire complex of lithiumcoordinated with the anions move together. However, in structure diffusion, the lithium ion transports itself by exchanging ions in its first solvation shell. NMR spectroscopic studies on LiTFSIdissolved in two different imidazolium ILs revealed that the lithium ions were coordinated with at least two or more TFSI ions in the system which lead to the formation of complexes with net effectivenegative charge[19]. These complexes give rise to the poor performance of the electrolyte as they resist the deposition of lithium on the cathode[18, 19]. However, a study by Borodin et. al[17] suggested that the vehicular motion mechanism only contributed to 30% of lithium diffusion and therefore the lithium diffusion can take place unhindered via structure diffusion mechanism. Although, it was observed that upon increasing the salt concentration in the solvent led to vehicular motion playing a more dominant role[19]. It has also been found that dispersing the solvent in polymer network with polar sites increases the structure diffusion of acetamide significantly, by reducing the solvation of anions around the lithium[19].

Recently, we have carried out a detailed study of diffusion mechanism of acetamide in a DES synthesized from acetamide and lithium perchlorate ($LiClO_4$). It was established that the long-range diffusion of acetamide is strongly hindered in the DES compared to molten acetamide[20]. Although this has been ascribed to the formation of H-bonds between the ionic species and acetamide molecules in the DES, a detailed analysis of the underlying mechanism of dynamics is not well understood. In this work, we use the explicit atomistic details accessible through MD simulations to address this issue. Further, the main focus of this work is to understand the diffusion of lithium ions in the DES. Subsequently, the role of the acetamide diffusion on lithium transport is also investigated. We also give a brief account of the neutron scattering experimental validation of the findings of MD simulation.

## 2. Materials and Methods

### 2.1 Molecular Dynamics Simulation protocol

An all-atom molecular dynamics (MD) simulation of acetamide with lithium perchlorate was carried out in the molar ratio of 78:22, to simulate the deep eutectic solvent (DES). The initial configuration of the system was prepared by randomly arranging 400 acetamide, 56 lithium and perchlorate ions in a cubic box, to account for the appropriate mole ratio. The CHARMM forcefield version 27[21] was used to parameterise acetamide, while for ions the parameter set was taken from the literatures[22, 23]. The chosen set of parameters for the DES has been already established to reproduce the experimental dynamical models quite accurately[20]. The initial configuration of the system was relaxed to equilibrium by carrying out the simulation in the NPT ensemble, using Langevin thermostat and barostat for 10 ns. The target pressure was set at 1 atm and independent simulations were carried out at 365 K. A production run 5 ns was carried out for the system in the same NPT ensemble, recording trajectory of all atoms at interval of 1 ps. Moreover shorter production runs were also carried out independently to record the trajectories at shorter time intervals of 0.01 ps. All simulations were carried out using DL_Poly_4[24].

### 2.2 Calculation of Correlation functions – MD simulations

The major advantage of MD simulations is the ability to obtain atomistic information from simulation trajectories that are inaccessible from experiments. The hydrogen-bonding (H-bonding) nature of the system can be characterised by the H-bond time correlation function which can be explicitly calculated from MD simulations. An H-bond between two molecules/ions can be characterised by geometric conditions, i.e., if the donor (oxygen in acetamide or perchlorate) and acceptor (amide nitrogen or lithium) are within a cutoff distance of $R_{cut}$. Further, when characterising H-bond between two molecular structures, a cutoff angle, ($\theta_{cut}$) is also considered between atoms connecting the H-bond[25]. The values of $R_{cut}$ for acetamide-acetamide and acetamide-ion pairs are obtained from the first minima of their respective pair distribution functions (PDFs). In the case of acetamide-acetamide and acetamide-anion pairs, the cutoff distance is taken as 4 Å, with cutoff angle of $30°$[20, 25-27]. In the case of acetamide-cation pairs, the $R_{cut}$ is chosen to be ~ 2.3 Å based on lithium-oxygen(of acetamide) PDF[20, 25]. The H-bond correlation functions, $C_h(t)$, are calculated from flags, $h_i(t)$, which take values of 1 if the $i^{th}$ pair is H-bonded and zero otherwise, with the following formula,

$$C_h(t) = \frac{1}{N_{pairs}} \sum_{i=1}^{N_{pairs}} \frac{\langle h_i(t+t_0)h_i(t_0) \rangle}{\langle h_i(t_0)^2 \rangle} \quad (1)$$

where, $N_{pairs}$ is the total number of pairs, $t_0$ is an arbitrary time origin, and the angular brackets denote average over these arbitrary time-origins.

It is also convenient to calculate various correlation functions that can be directly compared with the experimental data, among which, mean-squared displacement (MSD) and intermediate incoherent scattering function (IISF), are quantities useful in probing the dynamics of the system. The MSD of any molecule/atom are calculated from the MD simulation trajectories as,

$$\langle \delta r^2(t) \rangle = \frac{1}{N} \left\langle \sum_{n=1}^{N} [\mathbf{r}_n(t+t_0) - \mathbf{r}_n(t_0)]^2 \right\rangle \quad (2)$$

where, $\mathbf{r}_n(t)$ is the position of the $n^{th}$ molecule/atom and $N$ is their total number in the simulation box. The angular brackets indicate the same as indicated for eq. (1). While MSD is a good indicator of the typical timescale of relaxation process, its detailed profile can be obtained by modeling IISF, which can be calculated from MD simulation trajectories as follows,

$$I(Q,t) = \frac{1}{N} \left\langle \overline{\sum_{n=1}^{N} e^{i(\mathbf{Q}\cdot[\mathbf{r}_n(t_0+t)-\mathbf{r}_n(t_0)])}} \right\rangle \quad (3)$$

where $Q$ is the reciprocal space vector and the bar denotes average over all orientations in the reciprocal space. The values of $Q$ are restricted by the geometry of simulation box, given by $Q = 2n\pi/L$, where L is the edge length of the cubic box. The other symbols have the same meaning as mentioned earlier for eq. (1) and (2). Further, the calculation of the IISF also gives us the advantage to compare with incoherent scattering function obtained from neutron scattering experiments, through a time-Fourier transform[28].

*2.3 Neutron scattering experiments*

The deep eutectic solvent (DES) was synthesised from a solid mixture of acetamide and lithium perchlorate, in the mole fraction 78:22, which was kept heated to a temperature ~ 340 K for about an hour, until a clear solution was formed. The obtained DES remained in liquid state upon cooling down to room temperature (300 K). Quasielastic neutron scattering (QENS) experiments on the DES at 365 K were carried out at the FOCUS spectrometer in Paul Scherrer Institute (PSI), Switzerland. The spectrometer was used with incident neutron wavelength of 6 Å, which offers an energy resolution of ~ 45 μeV and a $Q$-range of 0.4 to 1.6 Å$^{-1}$. Standard vanadium was used to measure the resolution of the spectrometer.

The obtained neutron scattering spectra are dominated by incoherent scattering owing to the enormously large incoherent scattering cross-section of hydrogen atoms[28]. The incoherent scattering law, $S_{inc}(Q,E)$ is used to directly model the relaxation processes or dynamics in the system as it is related to the time-Fourier transform of the IISF. The experimental data is fitted to the convolution of the theoretical model scattering law with the instrumental resolution function, $R_{inst}(E)$,

$$S_{inc}^{epxt}(Q,E) = S_{inc}^{model}(Q,E) \otimes R_{inst}(E) \qquad (4)$$

## 3. Results and Discussion

Our method of analysis focuses on classifying the acetamide (ACM) molecules in different states based on their hydrogen-bonding (H-bonding) network with the neighbouring molecules and ions. Similar studies were carried out in order to understand the reorientational mechanism of acetamide in different DES by Das et al[25, 27]. The formation of different H-bond clusters and their lifetimes in these DES have also been studied in detail by Das et al[26]. Here we focus on studying H-bond structures in the system to enable us to understand their effects on translational mobility of acetamide molecules and transport of lithium ions in the DES.

A single acetamide molecule can act as both an H-bond donor (oxygen site) as well as an H-bond acceptor (amide hydrogen). The H-bond between the molecules and ions are defined using geometric conditions that have been described in section 2.2 in detail. While there is more than one acceptor site in each acetamide, we consider it to be a single possibility in our state definitions. The states of ACM are classified based on the following combinations with other acetamide/ions, represented by ions/acetamide attached to the acceptor site on left and ions/acetamide attached to the donor site on the right: (I) $ClO_4^-$ – ACM – $Li^+$, (II) ACM – ACM – $Li^+$ (III) none – ACM – $Li^+$ (IV) $ClO_4^-$ – ACM – ACM/none (V) ACM – ACM – ACM (VI) none – ACM – none. The keyword 'none' implies that the corresponding site is not H-bonded with any molecule/ion in the system. In particular, state (VI) refers to a free acetamide molecule. Similarly, in state (IV) both the possibilities of having another acetamide attached or no molecule/ion attached at the donor site are considered simultaneously. The state definitions were chosen based on considering different combinations that had significant population. Fig. 1 shows the time-averaged probability distribution of finding the acetamide in different states as categorised above. Snapshots of different states are also explicitly shown in Fig. 1; among the two possibilities in the case of state (IV), only $ClO_4^-$ – ACM – ACM is shown. It is evident that states (III) and (VI) are the most populated states, which indicates that apart from free acetamide, most of the acetamide molecules are preferentially bound to lithium ions.

In our previous work, we have established using neutron scattering and MD simulation that the dynamics of acetamide is slower in DES compared to molten acetamide[20]. The slowing down, particularly of the long range jump diffusion of acetamide, has been associated to the strong H-bonding of acetamide with the ions in the DES[20]. In order to gain deeper insight into the origin of the slow jump diffusion, we exploit the ability of MD simulation to demarcate acetamide molecules that are bound to ions from those that are free. Using the above category of states, we consider two most populated states, (III) and (VI), and labelled those acetamide molecules as bound and free if they remain at least 75 % of their time in their respective states for a duration of 1.1 ns. Considering the individual trajectories of bound and free acetamide molecules separately, their dynamics are analysed individually. This will be later compared to the dynamics of acetamide molecules in molten acetamide to understand the origin of their slow jump diffusion in DES.

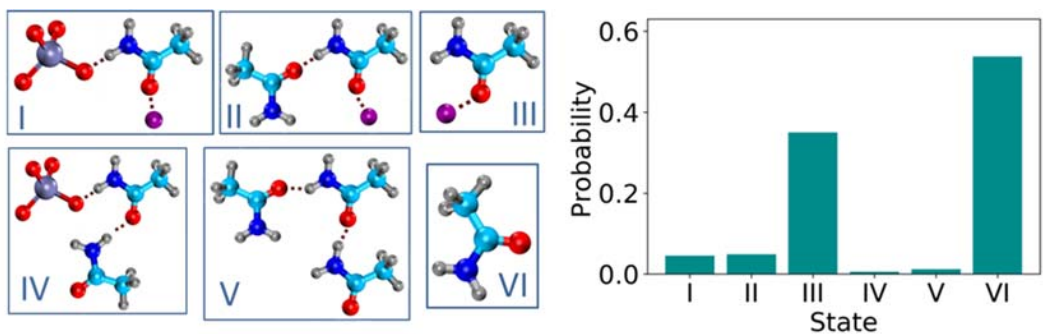

**Figure 1.** Time-averaged probability distribution of various states of acetamide as described in the text. Typical snapshots of each state are also shown (for state (IV), only the case of $ClO_4^-$ – ACM – ACM is shown) (hydrogen – gray; oxygen – red; carbon – cyan; nitrogen – blue; lithium – magenta; chlorine – purple).

### *3.1 Dynamics of bound and free acetamide in DES*

The mean-squared displacement (MSD) for the centre of mass (COM) of both bound and free acetamide molecules in the DES are calculated from their MD simulation trajectories (eq. 2) and shown in Fig. 2. The short time region ($t$< 0.1 ps) is the ballistic regime and it is fit with a quadratic time dependence indicated by the dotted line in Fig. 2. Clearly, in this regime, both the bound and free molecules behave similarly. Subsequently, after short subdiffusive regime, we have the diffusive motion ($t$> 50 ps), that has linear time dependence. The MSD in this regime is fit with Einstein's diffusion equation, $6Dt$, from which the diffusivities $D_f$ and $D_b$ are found to be $3.9 \times 10^{-6}$ cm$^2$/s and $1.3 \times 10^{-6}$ cm$^2$/s respectively. The linear fits are shown (for $t$> 50 ps) as solid lines for both free and bound acetamide cases in Fig. 2. Clearly, the bound acetamide molecules are at least three times slower compared to their free counterparts. Considering $p_f$(~ 0.4) and $p_b$(~ 0.6) as the fraction of molecules in the states (VI) and (III) respectively; as an immediate check, effective/average diffusivity of acetamide is calculated from ($p_f D_f + p_b D_b$), yielding a value of ~ $2.8 \times 10^{-6}$ cm$^2$/s, which is consistent with the diffusivity of acetamide from our earlier report[20].

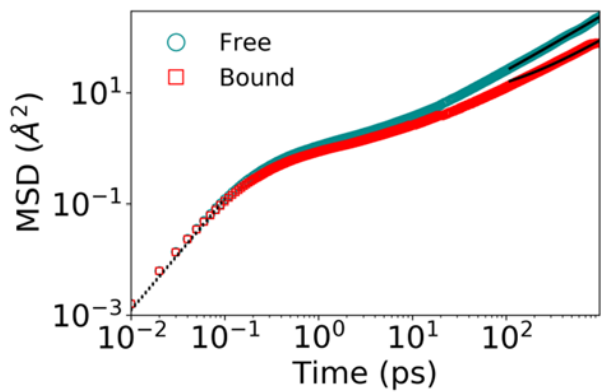

**Figure 2** The MSD of free and bound acetamide molecules in DES. At short times, the ballistic regime is indicated by the dotted lines. The linear diffusive regime is indicated by solid lines for both the cases.

Further, to proceed with the detailed analysis of the dynamics of bound and free acetamide molecules, their IISF, $I_b(Q,t)$ and $I_f(Q,t)$ were also calculated from their individual MD simulation trajectories using eq. (3). The diffusion model for acetamide in this DES has been established in our earlier comprehensive study of neutron scattering and MD simulation[20]. It was shown that the diffusion occurs via a combination of long-range jump diffusion and localised diffusion between jumps. The corresponding model function, including the ballistic component, for the IISF is given by,

$$I_{b/f}(Q,t) = \alpha(Q)\left[C_0(Q)e^{-\zeta_1 t} + (1-C_0(Q))e^{-(\zeta_1+\zeta_2)t}\right] + (1-\alpha(Q))e^{-(\sigma_v Q t)^2/2}$$
$$= \alpha(Q)C_0(Q)e^{-\zeta_1 t} + \alpha(Q)(1-C_0(Q))e^{-(\zeta_1+\zeta_2)t} + (1-\alpha(Q))e^{-(\sigma_v Q t)^2/2} \quad (5)$$

where, the subscript *b/f* indicates that this model will be used for both bound and free acetamide molecules in the system. The first term in the equation denotes the diffusive component, which is a combination of the global jump diffusion with the relaxation rate $\zeta_1$ and localised diffusion between jumps with the relaxation rate $\zeta_2$ and $C_0(Q)$ is the elastic incoherent structure factor of the localised diffusion motion. The second term corresponds to the ballistic motion of the molecules and $\sigma_v$ denotes the average thermal velocity of the molecules. The coefficient *α* gives the weight factor associated to the diffusive component in the IISF. The calculated values and fits for both $I_b(Q,t)$ and $I_f(Q,t)$ are shown Fig. 3 at a typical *Q*-value of 1.1 Å$^{-1}$; the individual components are also indicated in their respective plots; Comp. 1 & 2 (eq. 5) correspond to jump and localized diffusion processes respectively and Comp. 3 for the ballistic motion. The fits of IISF imply that the model used describes both bound and free acetamide quite satisfactorily. The characteristics of long range jump diffusion are obtained by analyzing the relaxation rate $\zeta_1$. Fig. 3(a) shows the variation of $\zeta_1$ with $Q^2$ for both bound and free acetamide molecules, clearly indicating that the former is significantly slower compared to the later. The relaxation rate $\zeta_1$ is modeled through the Singwi-Sjolander model of jump diffusion[29], given by,

$$\zeta_1(Q) = \frac{D_{b/f} Q^2}{1 + D_{b/f} Q^2 \tau_{b/f}} \quad (6)$$

where, $D_{b/f}$ and $\tau_{b/f}$ correspond to the jump diffusivity and mean residence time between jumps for bound and free acetamide molecules. Clearly, in the $Q \to 0$ limit, we recover the asymptotic Fickian diffusion with simple quadratic $Q$-dependence. The description based on eq. (6) for $\zeta_1$ is shown by the solid lines in Fig. 4(a), which indicate that the model employed is suitable. The jump diffusivities and mean residence time obtained from the fits are listed in Table 1. We can infer that in both the cases the diffusivities obtained from the linear fits of MSD (at long times) match well with the jump diffusivities, suggesting that the bound acetamide molecules are slower by a factor of two compared to the free molecules. It may also be noted that the mean residence time between jumps for the former is almost twice of that of the latter. It is observed that the jump diffusivity of free acetamide molecules in the DES is still smaller by a factor of 3 than acetamide molecules in the molten acetamide[20]. This difference might be associated to the effect of the chemical environment in DES, where the coulumbic interactions are much stronger compared to pure molten acetamide.

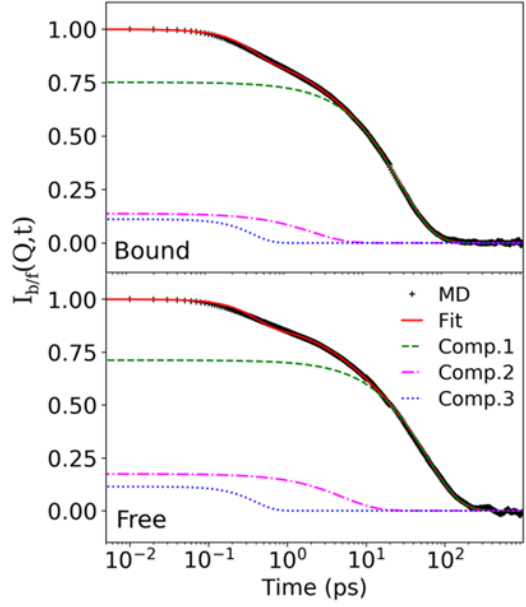

**Figure 3** IISF for bound (top) and free (bottom) acetamide molecules in the DES at $Q = 1.1$ Å$^{-1}$. The fitting based on eq. (5) and their individual components are also indicated in both the plots.

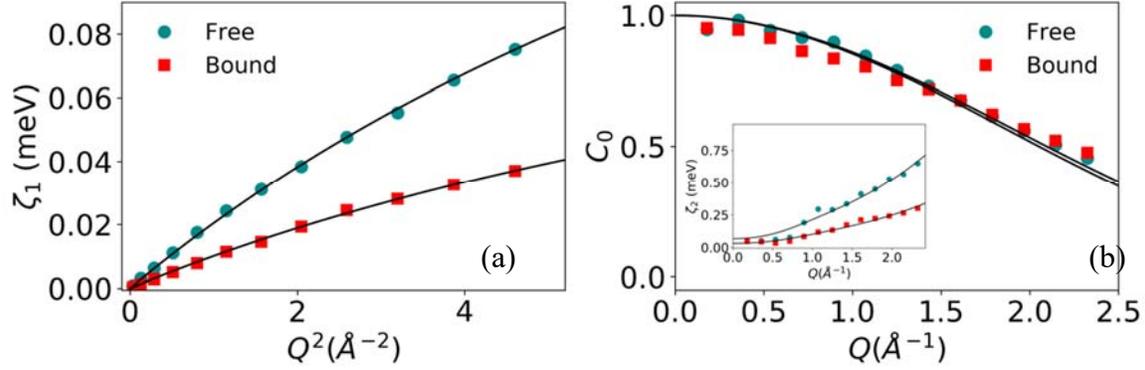

**Figure 4** (a) Variation of relaxation rate $\zeta_1$, associated to the jump diffusion process of free and bound acetamide molecules. The solid lines correspond to the fits based on eq. (6) using SS model. (b) Variation of EISF, $C_0(Q)$, associated to the localized diffusion process of both bound and free acetamide molecules. The solid lines are fits based on LTD model (see text) within spheres of exponentially distributed radii. (inset) Variation of relaxation rate $\zeta_2$, associated to the localized diffusion process for free and bound molecules along with their respective fits using LTD model.

**Table 1** List of dynamical parameters for bound and free acetamide molecules in DES at 365 K based on the model described by eq. (5). The numbers in the bracket denote the errors for the last decimal point.

| State | Jump diffusivity (cm²/s) | τ (ps) | $r_{avg}$(Å) | Localized diffusivity (cm²/s) |
| --- | --- | --- | --- | --- |
| **Bound** | $1.6 (2) \times 10^{-6}$ | 4.6 (1) | 0.8 (2) | $0.89 (2) \times 10^{-5}$ |
| **Free** | $3.3 (3) \times 10^{-6}$ | 2.2 (4) | 0.8 (1) | $1.88 (4) \times 10^{-5}$ |

The EISF, $C_0(Q)$, and the relaxation rate $\zeta_2$ are the parameters that describe the localized motion of acetamide between jumps. Fig. 4(b) shows the variation of these parameters for both bound and free acetamide molecules in the DES. As indicated by the EISF, the geometry of the localized dynamics is very similar in both the cases. However, behavior of $\zeta_2$ shows that the relaxation is much slower in the case of bound molecules. This dynamics has been shown to be well described by a modified version of localized translation diffusion (LTD) within spheres of exponentially distributed radii[20, 30]. The associated theoretical EISF for this model can be written as[20, 30],

$$C_0^{model}(Q) = 3\int_0^\infty dr \left[\frac{j_1(Qr)}{Qr}\right]^2 \frac{\exp(-r/r_{avg})}{r_{avg}} \tag{7}$$

where, $r_{avg}$ is the average radius of the exponential distribution. The fits based on above equation for the EISF, $C_0(Q)$, is shown in Fig. 4(b) by the solid lines and the obtained values of $r_{avg}$ are listed in Table 1. Evidently, the geometry of the local dynamics is not significantly different for bound and free acetamide molecules. The inset of Fig. 4(b) shows the variation of $\zeta_2$ with respect to $Q$ for free and bound molecules along with their respective fits as per the modified LTD model shown by the solid lines. The explicit details of the model and its analysis have been already described elaborately in our earlier report on transport mechanism of acetamide in the DES[20]. Table 1 lists the localized diffusion constants, obtained from the analysis of the relaxation rate $\zeta_2$. It is observed that, while the transient confinement radius of the molecules is the same in both the cases, the localized diffusion is at least two times faster in the case of free acetamide. In comparison with the pure molten acetamide, it is clear while the geometry of localized motion remains unchanged; the localized diffusion is much slower for both free and bound acetamide in the DES[20].

To summarize, we have established that there two kinds of acetamide molecules depending upon whether they're free or strongly bound to a lithium ion through an H-bond. The latter category of molecules show significantly slower jump diffusion compared to the average diffusivity in the system, which suggests these molecules play a crucial role in the slowing down of acetamide dynamics in DES compared to pure acetamide. Comparison of the experimental results from QENS is discussed at the end of this section. Prior to that, we shall describe the transport mechanism of lithium ions in the system and the role played by the bound acetamide diffusion in its transport.

### *3.2 Lithium diffusion in DES*

Lithium transport plays an important role in the application of DES as a liquid electrolyte for lithium ion batteries. The lithium H-bond with acetamide in the DES can be understood in more detail by studying the nature and dynamics of its solvation shell. The number of acetamide molecules in a solvation shell, denoted by $N_{ACM}$, for each lithium ion is calculated by counting the number acetamide molecules whenever it is H-bonded with lithium ion at each timestep. The average of number of acetamide molecules coordinating with each lithium ion is ~ 3.2. This is similar to the coordination numbers inLiTFSI in ethylene carbonate and imidazolium ionic liquids[18, 31]. The typical time-evolution of $N_{ACM}$ of a random lithium molecule in DES is shown in the inset of Fig. 5 (a). The complete probability distribution, P($N_{ACM}$), calculated from the MD simulation trajectories is also shown Fig. 5 (a). Evidently, 3 and 4 are the most dominant coordination of acetamide around each lithium ion. Snapshots of both configurations with 3 and 4 acetamide molecules H-bonded to a lithium ion is are shown in Fig. 5(b). It is now evident that the dynamics of bound acetamide molecule is significantly slower than acetamide in molten acetamide because, each of them is bound to other acetamide in the solvation shell of the lithium ions. Based on the nature of the solvation shell the diffusion of the lithium ions would be strongly governed by the dynamics of acetamide molecules in the system, in particular the bound acetamide molecules.

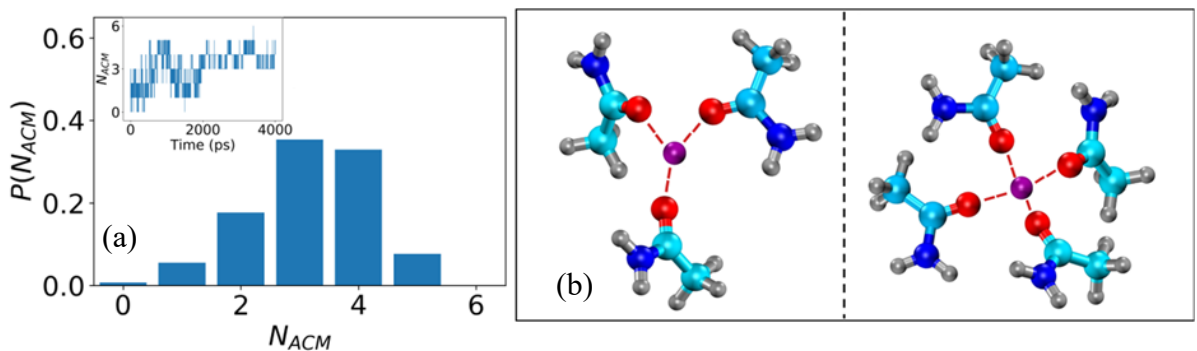

Figure 5 (a) Probability distribution of $N_{ACM}$ for the lithium ions in the DES. (inset) Time evolution of $N_{ACM}$ of a random lithium ion in the system. (b) MD simulation snapshots of lithium ion H-bonded with 3 and 4 acetamide molecules respectively. (hydrogen – gray; oxygen – red; carbon – cyan; nitrogen – blue; lithium – magenta; chlorine – purple).

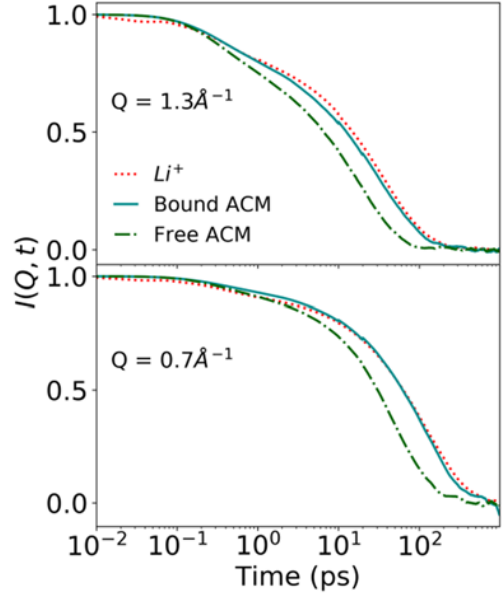

**Figure 6** IISF of the lithium ions, bound acetamide and free acetamide molecules in the DES calculated from MD simulation trajectories, at two typical $Q$-values.

In order to directly address this aspect of the problem, we calculated the IISF of the lithium ions from MD simulation and compare it with those of bound and free acetamide in the system. The comparative plots of all three IISF are shown in Fig. 6 at two typical $Q$-values. As it is clear from the plots, the IISF of lithium ions match remarkably well with that of bound acetamide, suggesting their dynamics are strongly correlated. To verify this quantitatively, we fit the IISF of lithium ions, with the same model (eq. (5)) as used earlier. Typical fit at $Q=1.1$ Å$^{-1}$ is shown in Fig. 7(a); the individual components of the fits are also shown. The quality of the fit suggests that this model works quite well for lithium diffusion too. The obtained relaxation rate, $\zeta_1$, corresponding to the jump diffusion of lithium ions is shown Fig. 7(b); the corresponding quantity for bound acetamide molecules is also shown by empty rectangular symbols for comparison. Firstly, we observe that the profile of the relaxation rates for lithium ions and bound acetamide are strikingly similar. Secondly, using the SS model (eq. 6) to fit the obtained $\zeta_1$ for lithium ions, we obtain a jump diffusivity of $1.5 \times 10^{-6}$ cm$^2$/s and a mean residence time of ~ 2 ps. Therefore, it is quantitatively evident that the jump diffusion of lithium and bound acetamide are strongly coupled. The inset of Fig. 7(b) shows the EISF of localized motion for lithium ions, along with the theoretical fit based on the modified LTD formalism (eq. 7), which yields an

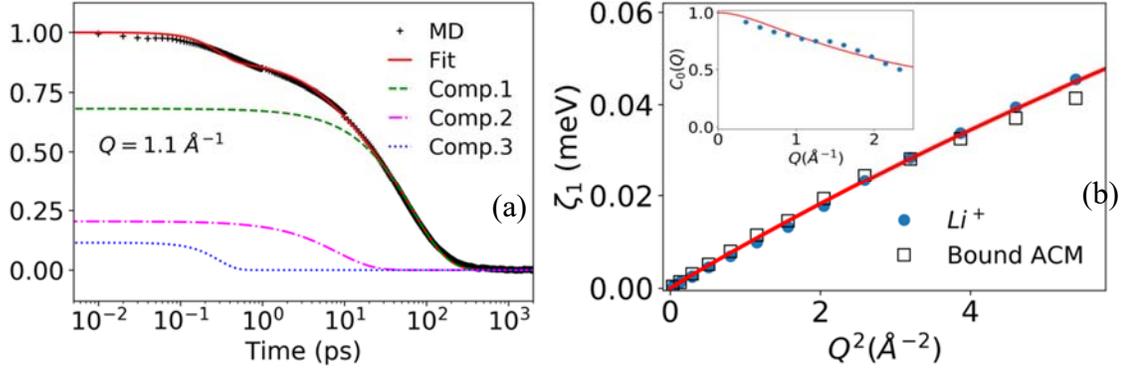

**Figure 7** (a) IISF of lithium ions calculated from MD simulations at $Q = 1.1$ Å$^{-1}$. Solid line is the fit based on eq. (5) and the respective components are denoted by dashed and dotted lines. (b) The variation of relaxation rate, $\zeta_1$, corresponding to jump diffusion of lithium ions and bound acetamide in the DES; the solid line indicates the fit based on SS model (eq. 6) for lithium ions. The inset shows the EISF, $C_0(Q)$, of the localized motion for lithium ions and its model function based on modified LTD given by eq. (7).

average radius of ~ 1 Å. Further, from the same model the localized diffusivity of lithium ion is also found to be around ~ $2.06 \times 10^{-5}$ cm$^2$/s. Unlike the case of jump diffusion process, the localized motion of lithium ions are much faster compared to bound acetamide, which is obvious due to much smaller mass and size of lithium compared to acetamide that plays a more dominant role in localized diffusion.

### 3.3 Transport mechanism of lithium ions: H-bond correlation

As mentioned earlier, the lithium transport in ionic liquid electrolytes is achieved by structure diffusion and vehicular diffusion mechanisms[17, 19]. The former one corresponds to diffusion of lithium through exchange of the molecules in its solvation shell[17] and in the latter case the lithium diffuses along with its first solvation shell[17]. From the analysis of IISF for the $Li$ ions, $I_{Li}(Q,t)$, it is found that the jump diffusion process of lithium is significantly slow and is almost equal to that of the bound acetamide. The jump diffusion is a long-range process and hence, will play a key role in the transport of ions in the DES media. Based on our observations, it is apparent that ionic transport of lithium is more likely to be dominated by the vehicular motion rather than structure diffusion process. However, to consolidate this claim further and obtain a quantitative estimate, we resort to analysis of the H-bond correlation function between

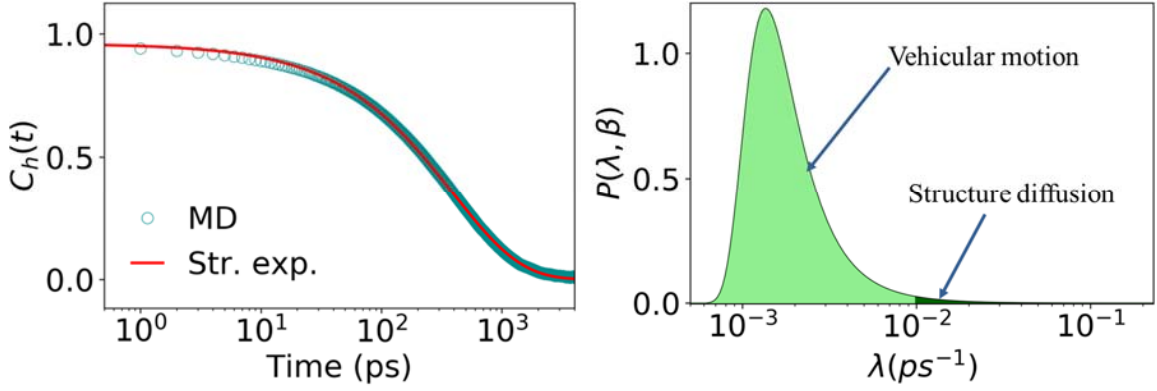

**Figure 8** (a) H-bond correlation function, $C_h(t)$, for acetamide-lithium pairs calculated from MD simulation (eq. 1). The fit based on stretched exponential (eq. 8) is also shown. (b) The distribution of relaxation rates for $C_h(t)$, calculated from eq. 9. The regions corresponding to vehicular motion and structure diffusion are shaded and indicated.

lithium ions and acetamide molecules which will shed light on the typical lifetime of the H-bond between the two. H-bond correlation, $C_h(t)$, for all the pairs of acetamide and lithium ions is calculated from MD simulation trajectories (eq. 1), as explained in section 2.2, and is shown Fig 8(a). The relaxation of H-bond dynamics is generally modeled using a stretched exponential function[18] owing to the possibility of multitude of relaxation rates in the system. Therefore we fit the correlation function with

$$C_h(t) = \exp\left[-\left(\lambda^* t\right)^\beta\right] \quad (8)$$

where $\lambda^*$ is the stretched relaxation rate and $\beta$ is the stretching exponent. The fitting of $C_h(t)$ as shown in Fig. 8 (a) yielded a value of $\beta \sim 0.78$, and $\lambda^* \sim 390$ ps. Clearly, the obtained fitting is quite good. It has been shown that the stretched exponential behavior can be ascribed to a distribution of relaxation rates in the system. Mathematically, it can be expressed as[32],

$$C_h(t) = \frac{1}{\lambda^*}\int_0^\infty P(\lambda,\beta)e^{-\lambda t}d\lambda \quad (9)$$

where, $P(\lambda, \beta)$ is the distribution of relaxation rates, $\lambda$, in the system. The distribution can be directly computed by evaluating the following integral for a given value of $\beta$ and $\lambda^{*\,32}$,

$$P(\lambda,\beta) = \frac{1}{\pi}\int_0^\infty du\,\exp\left[-u^\beta \cos(\pi\beta/2)\right]\cos\left[\frac{\lambda u}{\lambda^*} - u^\beta \sin(\pi\beta/2)\right] \quad (10)$$

The distribution of relaxation rates, calculated for the obtained value of fitting parameters is shown in Fig. 8 (b). The distribution spans over a wide range of relaxation rate indicating that the typical H-bond lifetimes range from a few nanoseconds to picoseconds.

In order to estimate the fraction of lithium undergoing vehicular motion and structure diffusion, we split distribution of timescales into two regions. Structure diffusion of lithium ions can be recognized based on two necessary criteria. Firstly, they take place by exchange of the acetamide molecule in its solvation shell, which happens only when their H-bond is broken. Secondly, the displacement of the ions will be of the order of the size of an acetamide molecule (~ 3 Å). Given the jump diffusivity of lithium ions calculated in the previous section, we can estimate the typical time taken for such a displacement of $\delta x$ to be $6D_{Li}/(\delta x^2)$, which in the present case, is found to be ~ 100 ps. Therefore, the fraction of ions involved in structure diffusion can be calculated by integrating the distribution of H-bond lifetimes which are less than 100 ps, corresponding to relaxation rates beyond $\lambda \sim 10^{-2}$ ps$^{-1}$. Fig. 8(b) shows the region of the beyond $10^{-2}$ ps$^{-1}$ in the distribution, shaded in dark-green labelled as structure diffusion. Upon integration the contribution of structure diffusion is found to be 10 %. The Fig 8(b) also shows the region of vehicular motion shaded in light green, which accounts to an overwhelming contribution of 90 % in the system. Hence, it is now evident that a large majority of lithium ions in the DES are transported through vehicular diffusion mechanism rather than structure diffusion.

*3.4 Neutron scattering –experimental validation*

Finally, we present a brief experimental corroboration for two dynamically distinct kinds of acetamide molecules - bound to lithium and free. Neutron scattering in any hydrogeneous media is dominated by incoherent scattering from hydrogen in the system. Incoherent quasielastic scattering gives us information about the stochastic dynamics in the system. Therefore, diffusion of acetamide can be directly observed from QENS data of the deep eutectic solvent. Unlike MD simulations, the contribution of bound and free acetamide cannot be individually observed in QENS experiments. However, we can obtain a combined scattering law for considering both. Let $p_f$ and $p_b$ be fraction of acetamide molecules in free and bound states respectively. The model scattering law for the entire system is written as,

$$S_{inc}^{model}(Q,E) = p_b S_b(Q,E) + p_f S_f(Q,E) \tag{11}$$

with $p_b + p_f = 1$. $S_b$ and $S_f$ are the scattering laws associated to bound and free acetamide molecules. The individual scattering laws for both the cases are given by a Fourier transform of the diffusive part of equation (5),

$$S_{b/f}(Q,E) = C_0(Q)L_j^{b/f}(\Gamma_j,E) + (1-C_0(Q))L_{j+loc}^{b/f}(\Gamma_{j+loc},E) \qquad (12)$$

where, $L_j$ and $L_{j+loc}$ are the Lorentzians corresponding to jump diffusion and jump+localised diffusion with $\Gamma_j$ and $\Gamma_{j+loc}$ as the respective half-width at half maximum (HWHM) and $C_0$ is the EISF of localised motion. Here we have ignored the ballistic component as it is significantly fast and will not contribute in the observable range of a neutron spectrometer. It should also be noted that while different Lorentzians are considered for bound and free acetamide molecules, the same EISF, $C_0(Q)$, is considered for both – as inferred from the results of MD simulation. Using eq. (12) and (11), the complete model scattering law can be expressed as,

$$S_{inc}^{model}(Q,E) = C_0(Q)\{p_b L_j^b(\Gamma_j,E) + (1-p_b)L_j^f(\Gamma_j,E)\} + (1-C_0(Q))L_{j+loc}^{b+f}(\Gamma_{j+loc},E) \qquad (13)$$

wherein we have combined jump + localised diffusion component of both the bound and free acetamide molecules. The Lorentzian $L_j^b$ and $L_j^f$ correspond to the jump diffusion of bound and free acetamide molecules respectively and the Lorentzian $L_{j+loc}^{b+f}$ corresponds to the combined jump + localised diffusion for both of them. As described in eq. (4), the experimental QENS spectra are fit by convoluting the eq. (13) with the resolution of the spectrometer. From the MD

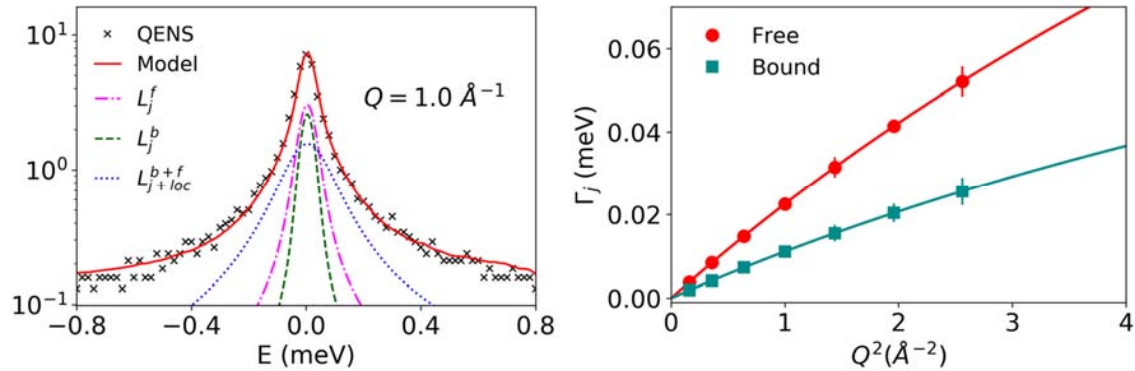

**Figure 9** (a) Typical fit of QENS spectra based on eq. (13), shown along with the three individual components. (b) HWHM of jump diffusion for bound and free acetamide molecules obtained from the fits of QENS spectra. The solid lines indicate their respective fits according to the Singwi-Sjolander jump diffusion model.

simulations, we have fixed the values of $p_b$ to be ~ 0.4. The fits of the QENS spectra at a typical $Q$-value of 1.0 Å$^{-1}$ is shown in Fig. 9 (a) along with all the three individual components.

Our primary interest is to validate the results of MD simulation for the jump diffusion mechanism of bound and free acetamide molecules in the system, which is characterised by the HWHM of jump diffusion, $\Gamma_j^{b/f}$. The obtained value of $\Gamma_j^{b/f}$ from the fits of eq. (13) are shown in Fig. 9 (b), it is clear that the bound acetamide molecules have significantly slower jump diffusion. The jump diffusion for free and bound molecules is modelled using the SS jump diffusion model described by eq. (6) and their respective fits are shown in Fig. 9 (b). The quality of fits indicates that the SS model description of jump diffusion is remarkably good for both the kinds of acetamide molecules in the system. In the case of bound acetamide, the model yielded a jump diffusivity of $1.8 \times 10^{-6}$ cm$^2$/s and mean residence time of ~ 4.2ps; while in the case of free, the jump diffusivity was found to be $3.7 \times 10^{-6}$ cm$^2$/s with average residence time of ~ 2 ps. Comparing the parameters obtained from experiment with those obtained from MD simulation (Table 1), it is apparent that the chosen model, indicating the presence of two dynamically distinct species of acetamide, sis consistent

## 4. Summary and Outlook

In this work, we have elucidated the underlying cause for the very slow long range diffusion of acetamide in the deep eutectic solvents (DES). It has been traced to the existence of two dominant species of acetamide; (i) H-bonded to lithium ions and (ii) completely free of any H-bonds. Analysis of their individual dynamics showed that the long-range jump diffusion of the former is significantly slower than the latter. This is also corroborated by the quasielastic neutron scattering results. As the viscosity of the solvent is coupled to long-range diffusion process, it is suggestive that the existence of such strong H-bonding with ionshinders the diffusion which could be the origin of the solvent's large viscosity.

It was found that the lithium ions were solvated in the DES, with its solvation shell predominantly made up of 3-4 acetamide molecules. Moreover, it was observed that the dynamics of acetamide played an important role in the transport of lithium ions in the DES. In particular, the jump diffusivity of lithium ions was found to be sameasthat of bound acetamide molecules which were H-bonded to it. This indicated that the vehicular motion might be the more favorable mechanism of lithium transport, in which the lithium ions move along with their solvation shells. This was rightly validated by analyzing the relaxation of H-bond correlation function between lithium and acetamide molecules in the system. The analysis also yielded a quantitative estimate, that 90% of the lithium ions in the DES were transported by the vehicular motion and only a meager 10% by the structure diffusion which is achieved by exchange of molecules in its solvation shell. This implies that lithium ions move along in a complex comprising of 3-4 acetamide moleculesattached around the ion. It is crucial to note that, unlike in the case of ionic liquid electrolytes, the vehicular motion in the DES will not be retarded in the presence of an externally applied electric field. This is because, as opposed to the case of ionic liquids, the complex of lithium ion (in DES) does not possess a net negative charge but rather, is neutral.However, the relatively slower transport of lithium ions by vehicular motion can probably be improved by dispersion of a polymer network with polar groups, which might enhance the structure diffusion of the ions in the system.

**Acknowledgement**

We are grateful to Dr. Jan Peter Embs from Paul Scherrer Institut (PSI), Switzerland for fruitful discussions.